\documentstyle[prc,aps,preprint,psfig]{revtex}
\begin{document}
\title{$e^+e^-$ pairs from $\pi^-$A reactions\thanks{Work 
supported by DFG, BMBF, and GSI Darmstadt}}
\author{M. Effenberger, E.L. Bratkovskaya, W. Cassing, and~U.~Mosel\\
Institut f\"ur Theoretische Physik, Universit\"at Giessen\\
Heinrich-Buff-Ring 16, D-35392 Giessen\\
UGI-99-01}
\maketitle
\begin{abstract}
We calculate dilepton production for the reactions $\pi^-$C and $\pi^-$Pb 
at 1.3 GeV 
within a semi-classical BUU transport model and compare our results to 
a previously published calculation. We show that a modified treatment of
the $\rho$-meson production and propagation
gives substantially different results. We, furthermore, discuss 
uncertainties related to the electromagnetic decay of the $\rho$-meson 
and the elementary $\pi^- N \to e^+ e^- X$ channels.
\end{abstract}

\section{Introduction}
The spectroscopy of vector mesons $(\rho, \omega, \phi)$ by their dileptonic
decay in finite or dense
nuclear matter is of great interest \cite{CB99} and new spectrometers are
currently being built \cite{HADES}. Whereas dileptons from nucleus-nucleus
collisions are complicated to interpret due to the complex dynamical evolution,
$e^+ e^-$ pairs from photon-nucleus, proton-nucleus or 
pion-nucleus reactions essentially probe
vector meson properties at normal nuclear matter density provided that appropriate 
cuts on the (low) momentum-spectrum of the dileptons are applied.
\par In Ref.~\cite{weidmann} dilepton production in pion-nucleus reactions 
has been calculated within the framework of a BUU transport model 
\cite{teis}. For the production and propagation of vector mesons a 
'perturbative' scheme was imposed where the perturbative particles 
were treated different from the non-perturbative ones. Especially the finite width of the $\rho$-meson was neglected in the production part and only taken into account for the dilepton spectrum by means of a formfactor. 
Meanwhile we have developed, starting from the very same transport 
model, a computer algorithm which incorporates the properties of 
perturbative particles in a dynamical way in line with our treatment of non-perturbative particles. Within this model we have 
calculated photoproduction of dileptons in nuclei in the energy range 
from 500 MeV to 2.2 GeV \cite{photo}. Since this model, that also contains a number of other improvements, gives different 
results for pion induced dilepton production than previously published \cite{weidmann} 
we want to discuss these differences in this article.

\section{The model}
For a complete description of the underlying model we refer to 
Ref.~\cite{photo}. Here we only briefly describe the main 
differences with respect to the earlier calculations:

\begin{itemize}
\item For the elementary meson-nucleon interaction we have meanwhile 
adop\-ted all resonance parameters from Manley {\it et 
al.}~\cite{manley} including some additional high-mass 
resonances. Especially the decay channel $R \to \Delta \rho$ is now included.
\item The finite widths of the $\rho$- and $\omega$-mesons are taken 
into account dynamically. In-medium changes of their spectral 
functions due to collisional broadening are treated analogously to our 
description of baryonic resonances \cite{abspaper,prodpaper}.
\item The production and absorption of $\rho$-mesons are now consistently
described within the resonance model of Manley {\it et 
al.}~\cite{manley}.
\item For the electromagnetic decay of the $\rho$-meson to $e^+e^-$ we 
use now a width proportional to $M^{-3}$, as resulting from vector 
meson dominance (VMD) \cite{svmd}, instead of one proportional to $M$ 
from extended VMD \cite{evmd}, with $M$ being the invariant mass of the 
$\rho$-meson. For our calculations this is more appropriate
since we neglect a direct coupling of the virtual photon and can not treat 
the resulting interference terms properly within a semi-classical transport
approach.
\end{itemize}
\section{Results}
\begin{figure}[h]
\centerline{
{\psfig{figure=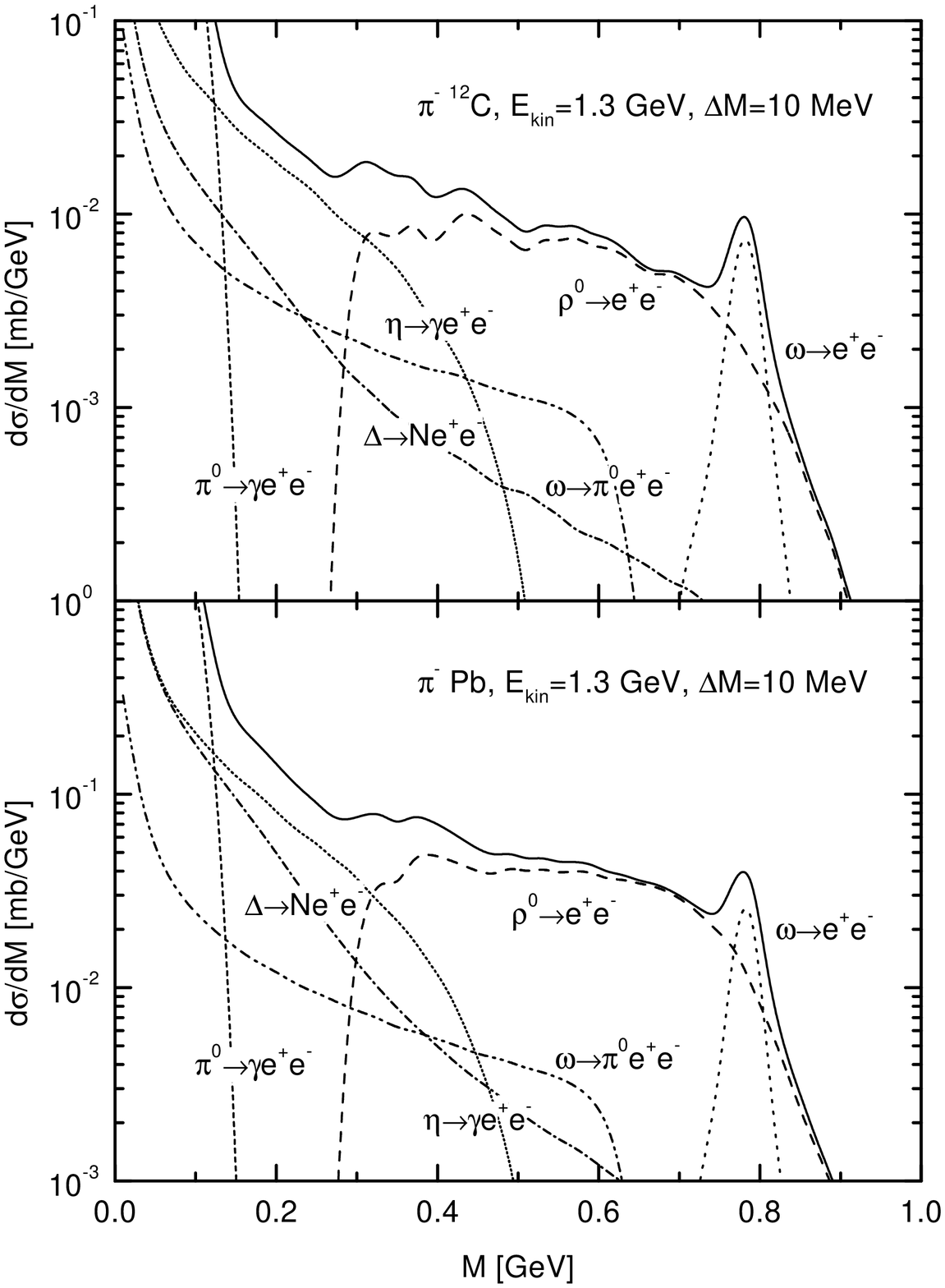,width=10cm}}}
\caption{The dilepton invariant mass spectrum for $\pi^-$C (upper part) and $\pi^-$Pb (lower part) at a 
kinetic energy of $E_{kin}=1.3$ GeV calculated without collisional 
broadening and with vacuum masses for the vector mesons employing a mass resolution $\Delta M=10\,{\rm MeV}$. Fluctuations in the curves are caused by low statistics.}
\label{fig1}
\end{figure}

In Fig.~\ref{fig1} we show the results of our calculations for 
$e^+e^-$-production in $\pi^-$C and $\pi^-$Pb reactions at a kinetic energy of 1.3 
GeV. Here neither collisional broadening nor an in-medium mass shift of 
the vector mesons are taken into account.
In the figure the various contributions to the total dilepton yield stemming
from ($\pi^0, \eta, \omega, \Delta$) Dalitz decays as well as from
vector meson decays ($\rho^0, \omega$) are displayed.
Compared to the previous 
calculations from Ref.~\cite{weidmann}, but also to those of 
Ref.~\cite{golubeva}, our calculations give results which
are up to an order of magnitude larger at intermediate invariant masses $M$ for both the light and heavy system. The contributions 
from the $\rho$-meson and the $\Delta$-resonance are very different in 
size and in shape. The $\rho$-meson contribution is shifted to lower 
energies and much broader. This is basically due to three reasons. 
Firstly, the modified dilepton decay width introduces a factor 
$(M_\rho/M)^4$ which, for example, at $M=0.5$ GeV gives a factor 5.6. 
Secondly, in our new calculations some of the higher-lying resonances, 
especially the $D_{35}(1930)$ and the $F_{37}(1950)$, decay strongly 
into the $\Delta \rho$-channel. These decays give predominantly 
low-mass $\rho$'s and lead to a stronger contribution of the $\Delta$-resonance. Thirdly, secondary pions can, especially through the 
$D_{13}(1520)$-resonance, more easily contribute to $\rho$-production
in the low mass tail. In the earlier calculations this was strongly 
suppressed because $\rho$'s could only be produced with their pole 
mass.
\par The deviations of the new calculations from the earlier ones are 
therefore mainly related to different descriptions of the elementary 
$\pi N \to e^+e^- X$ process for which neither experimental data nor a 
reliable theoretical prediction exist. In Fig.~\ref{fig2} we, therefore, show
the dilepton spectrum for elementary $\pi^-$p and $\pi^-$n collisions which enter our
calculations as input. The $\rho^0$-contribution on the neutron is very 
different from that on the proton. This is due to the fact that, because of isospin, on the neutron only the $\Delta \rho$-channel contributes while on the proton the $N \rho$-channel is dominant. 
The discontinuity of the spectrum at the two-pion mass
is caused by our neglect of off-shell $\rho$-mesons with invariant masses 
below the two-pion mass.   
\par However, it is questionable if the contributions coming from the $\Delta \rho$-decay of some resonances are realistic since in the analysis of Manley {\it et al.} \cite{manley} only data for exclusive one- and two-pion production were taken into account and the channel $\Delta \rho$ was only included in 
order to absorb inelasticity. One should note that the incoherent 
resonance contributions to the reaction $\pi^+ p \to p \pi^+ \pi^+ 
\pi^-$ via intermediate $\Delta^{++}\rho^0$-states already exceed 
the experimental data \cite{landolt} by about a factor of 2. In 
Fig.~\ref{fig3} (upper part) we, therefore, show the result of a calculation for 
which we replaced the $\Delta \rho$ decay by the channel $\Delta 
\sigma$ where the $\sigma$-meson parametrizes a scalar, isoscalar two-pion 
state with mass $M=0.8$ GeV and width $\Gamma=0.8$ GeV. 
This gives a reduction of the dilepton yield at 
intermediate masses by about a factor~3.
\par In Fig.~\ref{fig3} (upper part) we also show the result of a calculation 
where we used
an $e^+e^-$-width of the $\rho$-meson proportional to $M$ instead of
the more consistent $M^{-3}$. This also gives a result which is more than a factor of 2 
different for dilepton masses around 500 MeV. 
\par Apart from the uncertainties discussed above it is
questionable if our description of dilepton production in
elementary pion-nucleon collisions is valid since we neglect interference
terms between the different contributions as well as all processes that can not be 
described by a two-step process. There might, for example, come a large
contribution from so-called $\pi N$ bremsstrahlung where the dilepton couples 
to the incoming pion.
\par In view of all these uncertainties in the theoretical description of the
elementary cross section it is necessary that the inclusive cross sections
for dilepton production on the nucleon are measured. Until then the following
results for dilepton production on nuclei are only an -- although state of
the art -- 'educated guess'.
\par During the last two years the $D_{13}(1520)$-resonance has 
received  great interest in connection with medium-modifications of 
the $\rho$-meson \cite{post,brown2,brat98,brat99}. 
In our calculations this resonance 
contributes to the production of low-mass $\rho$-mesons as well as to 
their absorption. About 30\% of the $\rho$-mesons in our calculations 
are produced via an intermediate $D_{13}$-resonance. In Fig.~\ref{fig3} (upper part) 
we show the result of a calculation where we excluded the 
$D_{13}$-resonance. Here we get a slight enhancement of the dilepton 
yield because absorption through this resonance is even more important than 
production.
\par In Fig.~\ref{fig3} (lower part) we show the result of 
a calculation where we assumed 'dropping masses' for the $\rho$- and 
$\omega$-meson \cite{brown}. We find a reduction of the vector meson peak 
around 770 MeV by about a factor 2.
The enhancement of the dilepton yield for masses around 
600 MeV is quite small because we already started from a quite flat 
$\rho$-meson 
contribution due to our implementation of the $\pi^-$n channel and neglected medium-modifications of the 
$N\rho$-widths of the baryonic resonances.
Therefore the total cross section for elementary $\rho$-meson production 
remains unchanged.  
\par In Ref.~\cite{photo} we describe in full detail how we implement the 
collisional broadening of the $\rho$- and $\omega$-mesons in our 
transport calculations in a dynamical way. In Fig.~\ref{fig3} (lower part) we
show the result of a calculation in which we took into account collision
broadening in addition to the mass shift.    
One sees that the effect of collisional broadening is small.
\par In Fig.~\ref{fig3} (lower part) we also present the result of a calculation with
a momentum dependent potential \cite{Kond98rho} for the vector mesons 
instead of the constant
mass shift. This potential gives the previously used mass shift for $p=0$,
increases linearly with momentum and crosses zero for $p=1$ GeV; for details 
see Ref.~\cite{photo}. The result for the dilepton spectrum is quite close to the calculation
without medium modifcations because the vector mesons are produced 
with rather large momenta in pion-nucleon collisions.
\par In order to discriminate between these 'scenarios' of
in-medium modification it is helpful to look on the spectra
for different momenta of the dilepton pair. 
In Fig.~\ref{fig4} we show the
results of our calculations for four different momentum bins. 
For low momenta ($p<300$ MeV) the 'dropping mass' scenario leads to a complete disappearance
of the vector meson peak around 780 MeV because a large fraction of the 
$\omega$-mesons with small momenta decays inside the nucleus. With increasing
momentum the fraction of $\omega$-mesons decaying outside the nucleus 
increases and therefore the 'vacuum peak' becomes more pronounced in the
'dropping mass' scenarios. The calculation with a momentum dependent potential
is getting closer to the calculation without medium modifications for larger
momenta since the momentum dependent potential vanishes for $p=1$ GeV.
\par In our calculations we assume an isotropic production of the vector mesons
in the pion-nucleon center of mass system because there are only  
experimental data on the angular distribution for larger energies. The spectra
shown in Fig.~\ref{fig4} depend strongly on the angular distribution in the 
elementary production step since different angles in the pion-nucleon center of
mass system correspond to different momenta in the laboratory frame. However,
a different angular distribution would primarily rescale the spectra
but hardly influence the qualitative effects of the medium
modifications.   
\section{Summary}
\begin{figure}[h]
\centerline{
{\psfig{figure=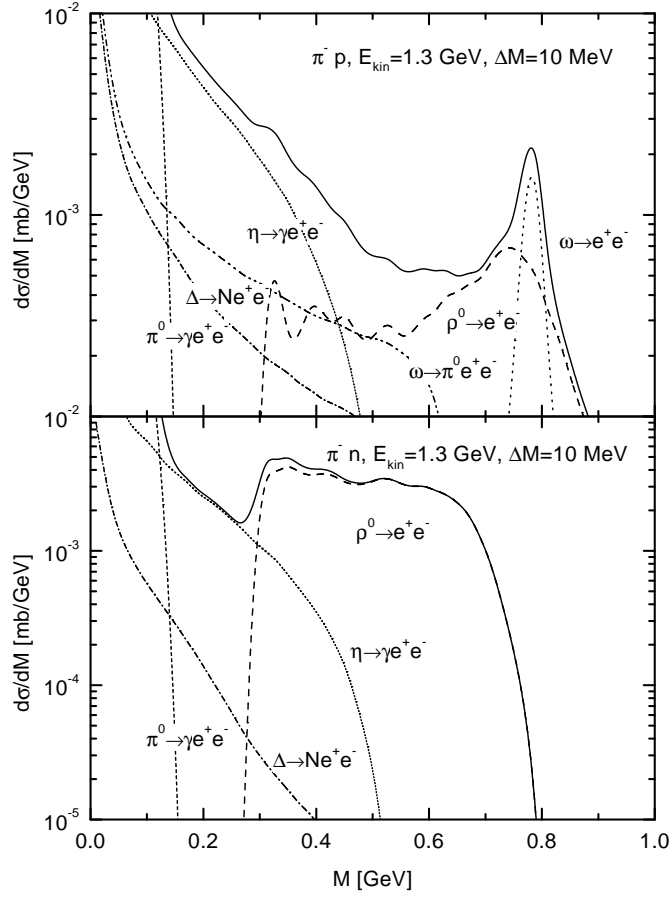,width=10cm}}}
\caption{The dilepton invariant mass spectrum for $\pi^-$p (upper part) and $\pi^-$n (lower part) at a 
kinetic energy of $E_{kin}=1.3$ GeV employing a mass resolution of $\Delta M=10 \, {\rm MeV}$.}
\label{fig2}
\end{figure}

\begin{figure}[t]
\centerline{
{\psfig{figure=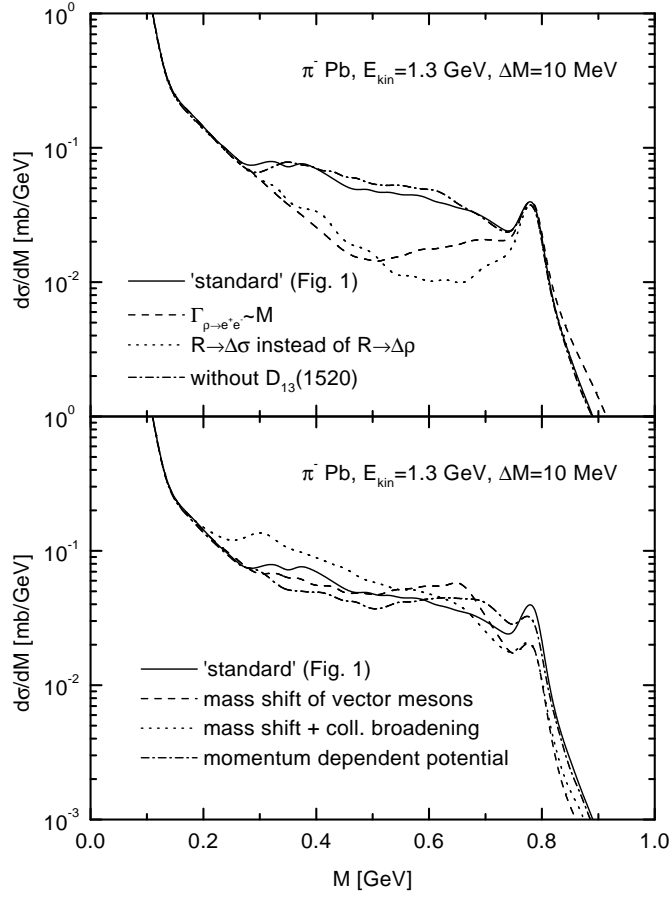,width=10cm}}}
\caption{Same as Fig.~\ref{fig1} for $\pi^-$Pb with different model assumptions (upper part, see text for a detailed explanation). The lower part shows the influence of 'dropping masses' and collisional broadening for the vector mesons.}
\label{fig3}
\end{figure}

\begin{figure}[t]
\centerline{
{\psfig{figure=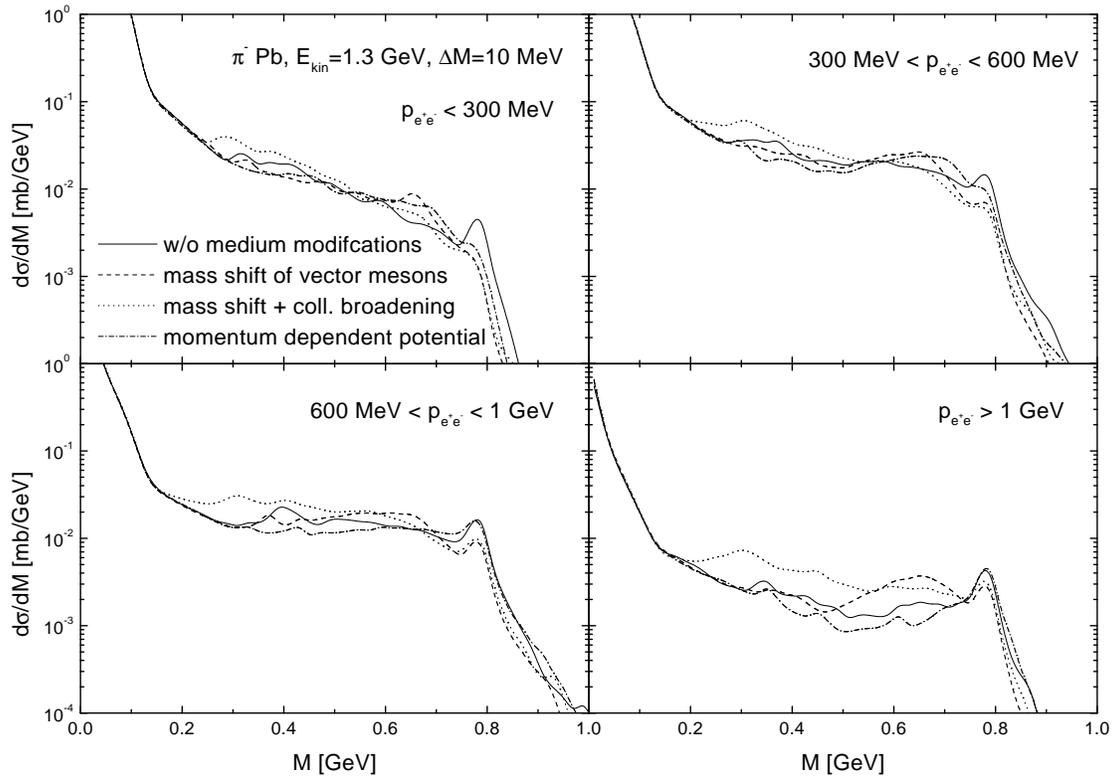,width=18cm}}}
\caption{Same as Fig.~\ref{fig3} for different laboratory momenta of the dilepton pair.}
\label{fig4}
\end{figure}

We have presented a calculation of dilepton production in $\pi^-$C and $\pi^-$Pb 
collisions at 1.3 GeV and compared our results to previously published 
calculations. We have discussed the uncertainties concerning the 
elementary $\pi N \to e^+e^- X$ cross section and want to stress the 
importance of an experimental measurement of the elementary process as 
prerequisite for reliable calculations in nuclei. The results shown in
Fig.~\ref{fig2}, for example, could be checked with the new spectrometer
HADES~\cite{HADES}, presently under construction at GSI. Here it would be 
quite desirable if measurements also for lower pion energies could be
performed since the contributions from secondary pions are important
for pion-nucleus collisions.
\par We have, furthermore, investigated the effects of different scenarios 
of in-medium
modifications for the vector mesons $\rho$ and $\omega$. Cuts on the momentum
of the dilepton pair might be helpful to distinguish between different scenarios.

\
\newpage

\end{document}